# Strain-Enhanced Mobility of Monolayer MoS₂

Isha M. Datye,[1] Alwin Daus,[1,2] Ryan W. Grady,[1] Kevin Brenner,[1,3] Sam Vaziri,[1] and Eric Pop[1,4,5,*]

[1]*Department of Electrical Engineering, Stanford University, Stanford, CA 94305, USA*
[2]*Present address: Chair for Electronic Devices, RWTH Aachen University, Aachen, 52074, Germany*
[3]*Present address: Department of Electrical and Computer Engineering, Southern Methodist University, Dallas, TX 75275, USA*
[4]*Department of Materials Science & Engineering, Stanford University, Stanford, CA 94305, USA*
[5]*Precourt Institute for Energy, Stanford University, Stanford, CA 94305, USA*

**ABSTRACT:**

Strain engineering is an important method for tuning the properties of semiconductors and has been used to improve the mobility of silicon transistors for several decades. Recently, theoretical studies have predicted that strain can also improve the mobility of two-dimensional (2D) semiconductors, e.g. by reducing intervalley scattering or lowering effective masses. Here, we experimentally show strain-enhanced electron mobility in monolayer MoS₂ transistors with uniaxial tensile strain, on flexible substrates. The on-state current and mobility are nearly doubled with tensile strain up to 0.7%, and devices return to their initial state after release of strain. We also show a gate-voltage-dependent gauge factor up to 200 for monolayer MoS₂, which is higher than previous values reported for sub-1 nm thin piezoresistive films. These results demonstrate the importance of strain engineering 2D semiconductors for performance enhancements in integrated circuits, or for applications such as flexible strain sensors.

Transition metal dichalcogenides (TMDs), a class of two-dimensional (2D) layered materials, have gained interest for electronic and optoelectronic devices due to their atomically thin nature and pristine interfaces that, in theory, lack dangling bonds.[1] Molybdenum disulfide (MoS₂) is a promising TMD because it can be synthesized in single layers,[2-3] it is relatively air-stable,[4] and its band gap is nearly twice that of silicon, which is advantageous for low-power transistors.[1] However, other electrical properties, including on-state current, mobility, and contact resistance of MoS₂ and other TMD devices must be improved for them to compete with or complement existing technologies based on silicon. Several techniques have been used to experimentally improve TMD-based transistors, such as contact engineering,[5-6] channel doping,[7-9] defect healing,[10] and interface engineering,[11] while strain engineering has been theoretically predicted as an additional method to improve electrical performance.[12-15]

Strain engineering was shown to improve the mobility of silicon metal oxide semiconductor field-effect transistors (MOSFETs) in the 1990s[16-17] and then commercialized with the 90 nm technology node.[18-20] In practice, electron mobility in nMOS silicon FETs is increased by uniaxial tensile strain from silicon nitride encapsulation layers.[19-20] In contrast, higher hole mobility in pMOS silicon FETs is achieved by uniaxial compressive strain imparted by selective growth of SiGe at the source and drain regions.[19-20] Reduced electron effective mass and scattering due to band splitting in the conduction band, and reduced hole effective mass due to band warping in the valence band lead to enhanced mobility with strain in silicon.[16,19-21] Experimental and theoretical studies have shown that strain can also modify the band structure and phonon dispersion of 2D semiconductors based on TMDs.[12-13,22-24] However, most strained TMD studies to date have focused on optical measurements (e.g. photoluminescence mapping of optical band gap changes with strain[24-26]), and less attention has been paid to strain effects on electron and hole mobility, despite the enhancement predicted theoretically.[12-15]

In this work, we study the effect of uniaxial tensile strain on the electrical performance of mono-layer $MoS_2$ transistors fabricated on flexible substrates. We find the mobility and on-state current are nearly doubled with ~0.7% applied strain, reverting to their initial values when the strain is removed. This represents the largest enhancement in TMD mobility with externally applied strain to date, revealing that strain engineering could be just as important as defect- and contact-engineering for enhancing the electrical performance of 2D transistors based on TMDs. We also show these devices can be used as strain sensors because they have voltage-dependent gauge factors up to ~200, larger than most conventional strain sensors based on bulk materials and most 2D-based strain sensors.

We fabricate $MoS_2$ transistors with local back-gates on free-standing polyethylene naphthalate (PEN), a flexible and transparent plastic substrate, as shown in **Figure 1**. The $MoS_2$ is grown by chemical vapor deposition (CVD) on separate $SiO_2$/Si substrates,[2] then transferred[27] onto the $Al_2O_3$ back-gate dielectric on PEN. We note that monolayer $MoS_2$ is also a flexible material and can withstand strains up to ~11%.[28] Detailed steps regarding the fabrication of these transistors are given in Supporting Information Section 1. We apply strain to our $MoS_2$ transistors using a two-point bending apparatus, by controlling the distance between the two ends of the substrate, as shown in **Figure 2a**. The strain can be estimated as $\varepsilon = \tau/(2R)$, where $\tau$ = 125 μm is the thickness of the PEN and $R$ is the radius of curvature of the bent substrate (see **Figure 2b**).[29] The strain applied to the $MoS_2$ is confirmed by Raman and photoluminescence (PL) spectroscopy, by monitoring the position of the in-plane E' peak at ~384 cm[-1] and the A exciton peak at ~1.8 eV, respectively. We note that we can only monitor *relative* changes in $MoS_2$ strain with bending, i.e. we cannot be sure if the transistor has built-in tensile or compressive strain from the transfer and fabrication process, because the Al₂O₃ gate dielectric can cause peak shifts in the MoS₂ Raman spectra.[30] Thus, all strain measurements below are reported relative to the as-fabricated devices in their flat, unbent state.

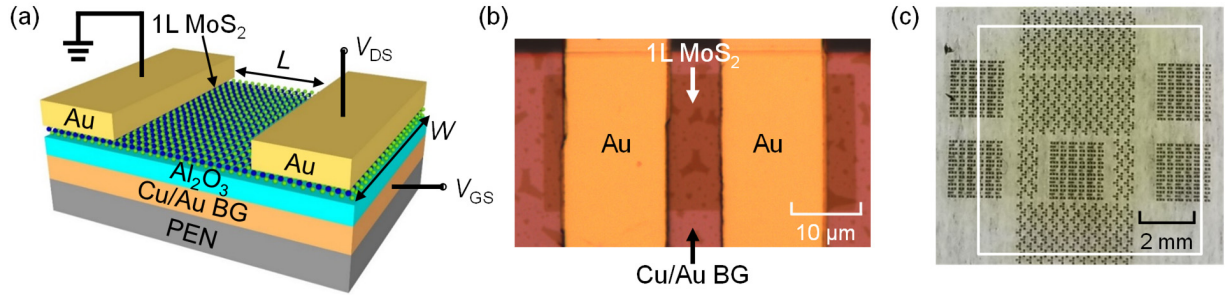

**Figure 1.** (a) Schematic of back-gated (BG) monolayer (1L) MoS₂ transistor on a polyethylene naphthalate (PEN) flexible substrate, with Au source and drain contacts, and an Al₂O₃ gate dielectric ~20 nm thick. (b) Top-view optical image of 1L MoS₂ device with length $L \approx 8$ μm and width $W \approx 20$ μm. A few small triangular bilayer regions are visible, but these do not bridge the transistor channel and we do not expect them to significantly affect the electrical performance of the monolayer MoS₂.[31] (c) Picture of sample (on a cleanroom wipe) taken after fabrication was completed, with the green region within the white box corresponding to the transferred MoS₂ film.

**Figure 2c** shows the Raman spectra of the MoS₂ device from **Figure 1b** without applied strain (blue) and with ~0.7% tensile strain (red). We perform Raman measurements on four locations across the device channel with similar results, but only include one representative spectrum at each strain level here for clarity. The E' peak clearly redshifts with ~0.7% applied tensile strain, and the average positions of the E' peaks without and with applied strain are $384.6 \pm 0.3$ cm⁻¹ and $383.0 \pm 0.2$ cm⁻¹, respectively. This corresponds to a peak shift of ~2.3 $\pm$ 0.2 cm⁻¹/% strain, which is comparable to that of other studies.[22,32-33] The cyan curve, representing the measurement after strain is released, matches very well with the initial 0% (blue) curve, indicating that the effects of strain are reversible. Figure S2 of Supporting Information Section 2 includes Raman peak position data for all devices measured and a short discussion of the smaller A₁' peak shifts.

**Figure 2d** displays the PL spectra of the MoS₂ device from **Figure 1b** at 0% (blue), 0.4% (magenta), 0.6% (red), and back to 0% strain (cyan). As expected, the A exciton redshifts with tensile strain because of a decrease in the direct, optical band gap at the K point (see **Figure 2e**).[24,26,34] The A exciton peak positions at 0%, 0.4%, 0.6%, and back to 0% strain are $1.810 \pm 0.006$ eV, $1.790 \pm 0.004$ eV, $1.772 \pm 0.004$ eV, and $1.811 \pm 0.005$ eV, respectively, averaged over four measurements in the device channel. Therefore, the shift of the A exciton peak is ~63 $\pm$ 10 meV/% strain. This is similar to that of other experimental studies demonstrating PL of MoS₂ with strain.[24,26,34] The PL peak position data for all devices measured are included in Supporting Information Figure S3b.

Next, we perform electrical measurements of our devices as a function of strain. Our set-up enables direct probing of transistors under strain (see **Figure 2a**) inside a vacuum probe station at ~2 × 10⁻⁵ torr pressure. **Figure 3a** displays drain current vs. gate voltage ($I_D$-$V_{GS}$) measurements of the MoS₂ transistor shown in **Figure 1b** with applied tensile strain from 0% to 0.7%, and back to 0%. The on-state current rises with increasing strain and then returns to the initial (unstrained) level after strain release (cyan curve in **Figure 3a**). Thus, strain does not have a permanent effect on the device characteristics. The gate leakage currents remain low (<1.2 nA) across all $V_{GS}$ and strain levels (see Supporting Information Figure S4). The drain current vs. drain voltage ($I_D$-$V_{DS}$) measurements of the device at 0% (solid lines) and 0.7% strain (dotted lines) are displayed in **Figure 3b** at several $V_{GS}$ values. As before, we observe $I_D$ increasing with strain; for example, at $V_{GS}$ = 7 V and $V_{DS}$ = 5 V, the current doubles from ~6 μA/μm to ~12 μA/μm at 0.7% strain.

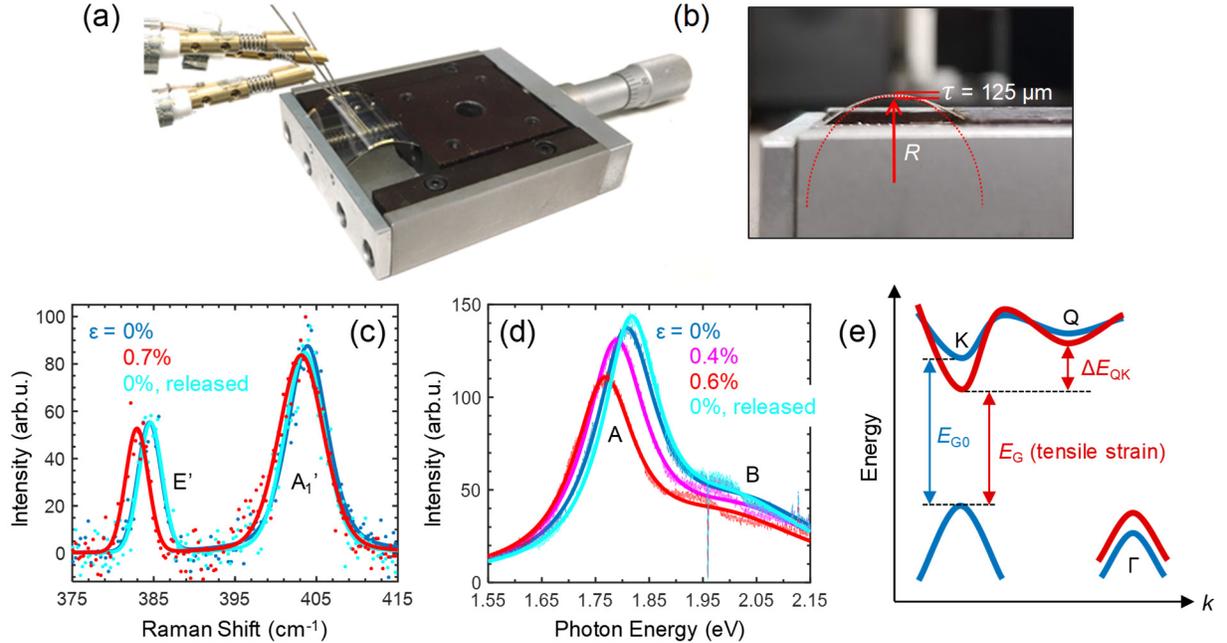

**Figure 2.** (a) Composite image of bending apparatus for applying tensile strain to the flexible substrates, illustrating the three probes (source, gate, drain) used for electrical measurements. (b) Image of bent substrate which is used to estimate the applied strain $\varepsilon = \tau/(2R)$, where $\tau$ is the thickness of the substrate and $R$ is the radius of curvature. (c) Raman spectra of the device from Figure 1b at different applied strains. The data (symbols) are fit to a superposition of Lorentzian and Gaussian peaks (solid curves). (d) Photoluminescence (PL) spectra of the same device, with the solid line fit to the data as the sum of two Lorentzians, for the A (at ~1.8 eV) and B exciton (at ~2.0 eV).[35-36] The A and B excitons are illustrated in the Figure S3a schematic. The line at ~1.96 eV is due to a Raman notch filter present in the optical path of the measurement setup. All Raman and PL measurements were taken in air, with laser wavelength 532 nm. (e) Simplified band structure of monolayer MoS₂, not to scale. With uniaxial tensile strain, the direct gap at the $K$ point decreases and the energy separation between $K$ and $Q$ conduction band valleys increases.[12-13,24,26] The valence band at the $\Gamma$ point also rises, leading to a narrowing of the K-$\Gamma$ indirect gap[15,26] and the decreased PL intensity observed with strain in Figure 2d.

We estimate the field-effect mobility from the $I_D$ vs. $V_{GS}$ curves in the linear operation regime, as $\mu_{FE} = (\partial I_D / \partial V_{GS}) L / (W C_{ox} V_{DS})$, where $C_{ox} \approx 312$ nF/cm$^2$ is the capacitance per unit area of the gate oxide, extracted from capacitance-voltage measurements of Au-Al$_2$O$_3$-Au structures on the same sample as the MoS$_2$ transistors, with and without strain (see Supporting Information Section 4). We note the threshold voltage ($V_T$) also changes with strain (see **Figure 3a** and Supporting Information Figure S11), suggesting that the electron density increases with the decreasing band gap under tensile strain.[33,37] To account for shifts in $V_T$, we estimate $\mu_{FE}$ at the same carrier density $n \approx 1.1 \times 10^{13}$ cm$^{-2}$, where $n = (C_{ox}/q)(V_{GS} - V_T - V_{DS}/2)$. We estimate $V_T$ using the linear extrapolation method, which uses the voltage intercept of a line fit to the $I_D$ vs. $V_{GS}$ data near the maximum transconductance $g_m = \partial I_D / \partial V_{GS}$.[38] The transistor has $\mu_{FE} \approx 5.3$ cm$^2$V$^{-1}$s$^{-1}$ with 0% applied strain and $\mu_{FE} \approx 10.8$ cm$^2$V$^{-1}$s$^{-1}$ with 0.7% applied tensile strain. Therefore, we achieve a ~2× improvement in $\mu_{FE}$ at 0.7% tensile strain for this device, which is essentially the same as the increase in drive current observed in **Figure 3b**.

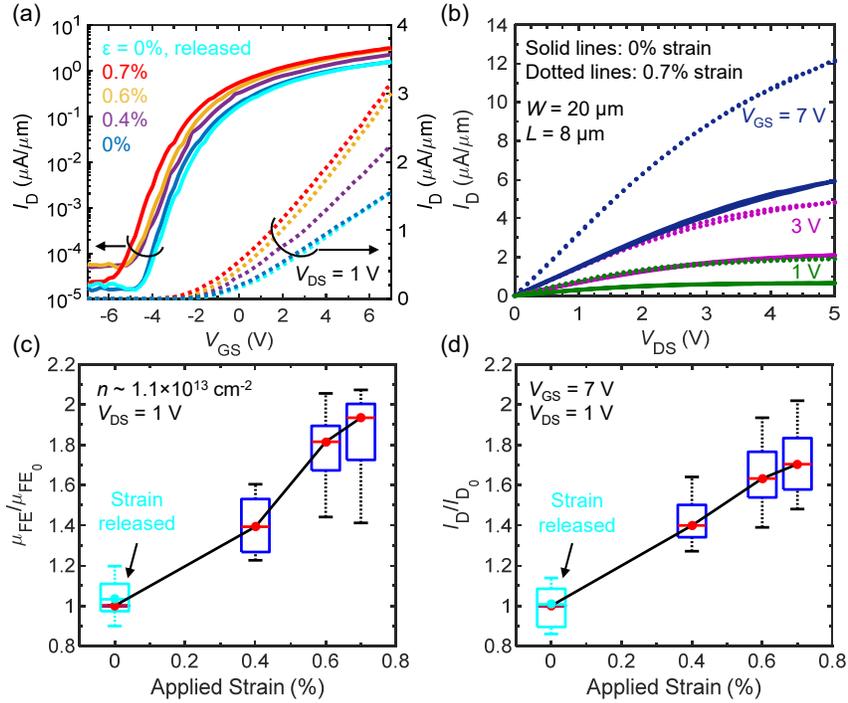

**Figure 3.** (a) Transfer characteristics ($I_D$ vs. $V_{GS}$) of the device from **Figure 1b** ($W = 20$ μm, $L = 8$ μm) at $V_{DS} = 1$ V and different levels of applied tensile strain. Solid lines correspond to data plotted on a log scale (left $y$-axis), and dashed lines correspond to the same data plotted on a linear scale (right $y$-axis). (b) Output voltage characteristics ($I_D$ vs. $V_{DS}$) of the same device at 0% (solid) and 0.7% (dotted) applied tensile strain, for $V_{GS} = 1$ V (green), 3V (magenta), and 7 V (dark blue). (c) Field-effect mobility ($\mu_{FE}$) normalized to the initial (unstrained) values for 8 devices as a function of applied strain, with the box plots showing the median across devices (red points), first and third quartiles (blue box), and maximum and minimum (top and bottom horizontal lines, respectively). The cyan box plot corresponds to the measurement after strain is released. (d) $I_D$ normalized to the initial (unstrained) values at different levels of strain, with the box plots again showing the distribution of values across all devices. $I_D$ values were extracted with applied voltages $V_{GS} = 7$ V and $V_{DS} = 1$ V. The cyan box plot again shows the measurement after strain is released.

We measure 7 other transistors with lengths from $L = 2$ μm to 15 μm on the same substrate and perform electrical measurements with applied strain. To account for $V_T$ variation, we extract $\mu_{FE}$ at the same carrier density ($n \sim 1.1 \times 10^{13}$ cm$^{-2}$) and $I_D$ at the same $V_{GS} = 7$ V and $V_{DS} = 1$ V, plotting the data for all devices in **Figure 3c-d**. These show $\mu_{FE}$ and $I_D$ normalized to their initial values without applied strain, with the data from all devices summarized by box plots. (Supporting Information Figure S6 shows the magnitudes of $\mu_{FE}$ and $I_D$ for each device, without normalizing them.) On average, $\mu_{FE}$ increases by a factor of $1.85 \pm 0.23$, and $I_D$ increases by a factor of $1.76 \pm 0.18$ with 0.7% applied tensile strain, compared to the initial values without applied strain. For the $I_D$ comparison, we note that the threshold voltage variation is at most $\delta V_T \approx 1$ V, which is significantly smaller than the overdrive voltage $V_{GS} - V_T \approx 8$ V. Therefore, the ~12.5% variation in electron density with $\delta V_T$ cannot account for the nearly 2× increase in $I_D$, which must come directly from the tensile strain applied.

Supporting Information Section 5 displays additional strain-dependent data as a function of channel length and at lower carrier density, with results largely consistent with the data presented in **Figure 3**. We also explored strain levels in excess of 0.7%, however we generally observed a degradation of electrical device performance in such cases (see Supporting Information Section 6). We attribute this degradation to either worsened adhesion of the contact metal to the MoS$_2$ or to cracking of the metal lines. Gate dielectric damage could be ruled out because gate leakage currents remained the same at higher strain levels; "slippage" of the MoS$_2$ along the substrate was also ruled out because we observed the expected shift of the E' Raman peak at higher strain levels (Supporting Information Figure S12). In future work, higher strains could be achievable with metals that are more ductile and with gate oxides like HfO$_2$, which has a lower Young's modulus than Al$_2$O$_3$.[39-40]

The improvements in current and mobility of our devices are expected to result from changes in the band structure with strain, as illustrated in **Figure 2d-e**. The direct band gap of monolayer MoS$_2$ at the K-point decreases with tensile strain, seen as a redshift of the A exciton in **Figure 2d**. The next-lowest valley in the conduction band is at the Q-point (approximately halfway along the T line between the Γ and K points[41-43]), and the energy separation between the Q and K valleys ($\Delta E_{QK}$) has been predicted to increase when tensile strain is applied to MoS$_2$ (see **Figure 2e**), resulting in less electron intervalley scattering and therefore improved mobility.[12-14,41] ($\Delta E_{QK}$ for monolayer MoS$_2$ encased in quartz and WS$_2$ was experimentally estimated to be ~110 meV,[44] and theoretical predictions are in the ~50-270 meV range for unstrained monolayer MoS$_2$, depending on the simulation approach used.[12-14,41,45-46] Monolayer TMDs such as WS$_2$, WSe$_2$, MoSe$_2$, and MoTe$_2$ have smaller $\Delta E_{QK}$, thus one may expect a larger mobility improvement with strain in transistors based on these materials.[13,46-47])

Tensile strain is also expected to change the curvatures of the conduction band valleys, leading to decreased electron effective mass.[14-15,48-49] This is similar to the reduced electron effective mass with strain in silicon nMOS transistors, which leads to increased mobility.[20-21] Applying tensile strain to 2D transistors has also been suggested to lower Schottky barriers at the source and drain contacts,[33,50-51] potentially leading to lower contact resistance. However, because our devices have relatively long channels and our improvements in $\mu_{FE}$ and $I_D$ do not depend on channel length (see Supporting Information Figure S7), we expect the contribution of contact resistance in our measurements to be relatively small.[52] Thus, we believe that the electrical performance improvements observed here are mostly related to electronic transport in the $MoS_2$ channel, i.e. lower intervalley scattering and effective mass.

We also consider the interaction of strain with defects in our $MoS_2$ channels. Electron transport in our $MoS_2$ channels is likely to occur in part by band-like transport, which is limited by scattering with phonons (e.g. intervalley[12-13,41]), defects, or impurities – and in part by hopping-like transport between defect trap states.[53] When strain is applied, the former benefits from lowering of the phonon-assisted intervalley scattering rate.[12-13,41] On the other hand, tensile strain could also influence the hopping-like transport, as suggested by recent results which saw improvements in carrier lifetime with tensile strain, due to weaker trapping of charge carriers.[54] In other words, tensile strain is expected to benefit the electron mobility for both "pristine", phonon-limited $MoS_2$ samples, and for lower-quality, "defective" $MoS_2$ samples.

The large change in resistance with strain of our $MoS_2$ transistors indicates that these devices could be useful for strain sensors. The figure of merit used to characterize strain sensors is the gauge factor (GF), defined as $(\Delta R/R_0)/\Delta \varepsilon$, where $\Delta R = |R_\varepsilon - R_0|$ is the change in resistance between $\varepsilon$ and 0% strain, and $R_0$ is the initial resistance with 0% applied strain.[55] Metals are often used in commercial strain gauges due to their ease of fabrication, but they typically have relatively low GF < 50.[56] Silicon strain sensors have GF ~ 30-50 for polysilicon[57-58] and up to 200 for single-crystal silicon.[59] 2D materials like $MoS_2$ and other TMDs are predicted to have large GFs because, similar to silicon and germanium,[60] they are piezoresistive. In addition, they can withstand much higher strains than conventional bulk materials, making them attractive for flexible electronics.[28]

**Figure 4a** shows the resistance ($R = V_{DS}/I_D$) vs. $V_{GS}$ curves for the device in **Figure 1b** from 0% to 0.7% applied tensile strain. **Figure 4b** illustrates $\Delta R/R_0$ vs. strain for several $V_{GS}$ = -4.4 V, 0 V, 3 V, and 7 V. We find the largest change in resistance at $V_{GS}$ = -4.4 V, which corresponds to the subthreshold region of the transistor (see **Figure 3a**). Fitting a dashed line to the $\Delta R/R_0$ vs. strain at this $V_{GS}$ yields an average GF ≈ 150. **Figure 4c** shows the calculated GF as a function of $V_{GS}$ for each strain level in **Figure 4a**. We observe a peak in GF at all strain levels around $V_{GS}$ = -4.4 V, with the maximum GF reaching ~200 for 0.4% tensile strain in this device. (We performed similar measurements on 7 other devices with channel lengths between 2 and 15 μm, and found an average maximum GF = 200 ± 45.) The GF displays a stronger gate dependence below and near threshold ($V_{GS}$ < 0) where current is limited by hopping-like transport between defect trap states,[53] recently shown to more weakly trap carriers when tensile strain is applied.[54] The GF is nearly constant in the linear transistor regime ($V_{GS}$ > 0, also see Figure 3a) where the band-like transport and mobility dominate.

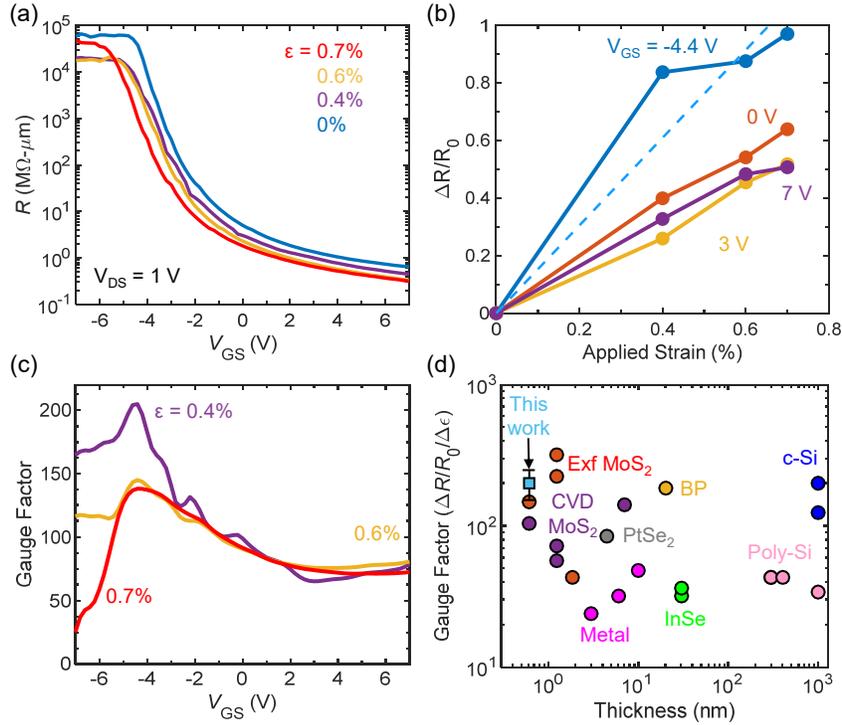

**Figure 4.** (a) Resistance ($R$) vs. $V_{GS}$ curves of the device in **Figure 1b** at $V_{DS}$ = 1 V at different levels of applied strain (ε). (b) $\Delta R/R_0$ vs. strain at different gate voltages for the curves in (a). (c) Gauge factor (GF = $\Delta R/R_0/\Delta \varepsilon$) vs. $V_{GS}$ for the different levels of strain. $\Delta \varepsilon$ is always calculated with respect to 0% strain. The GF below approximately -5 V is uncertain when the measurable $I_D$ minimum is reached (see Figure 3a), suggesting that the peak GF obtained could be higher at lower voltages. (d) GF vs. thickness for our best devices (shown as the blue square with error bars) in addition to various materials found in literature, including CVD-grown MoS₂, exfoliated MoS₂, indium selenide (InSe), platinum selenide (PtSe₂), black phosphorus (BP), polycrystalline Si (poly-Si), crystalline Si (c-Si), and thin metal films. We note that the c-Si values are for bulk Si with thicknesses likely greater than 1 μm.

**Figure 4d** compares the best GF values obtained in this work to those found in literature for MoS₂, other 2D materials, silicon, and metals (of various thicknesses). Our GF for monolayer CVD-grown MoS₂ is higher than the best GFs for monolayer and trilayer exfoliated MoS₂,[61] CVD-grown MoS₂,[62-64] other 2D materials (BP, InSe, and PtSe₂),[65-67] polysilicon,[57-58] and thin metal films.[56] Comparable or slightly higher GFs have been found for bulk crystalline silicon[59,68] and bilayer exfoliated MoS₂,[33,61] respectively. However, CVD-grown $MoS_2$ is easier to integrate and more promising than bulk silicon or exfoliated $MoS_2$ for large-area flexible and transparent sensors. We note that **Figure 4d** only displays GF calculated as $(\Delta R/R_0)/\Delta \varepsilon$, though some studies calculate GF as $(\Delta I/I_0)/\Delta \varepsilon$, where $I$ is current, which artificially yields much larger GF. For example, using the latter definition with current instead of resistance, the maximum GF achieved by our devices would be ~5000 instead of ~200.

In summary, we studied the mobility enhancement of CVD-grown monolayer $MoS_2$ transistors with tensile strain, by bending devices on flexible PEN substrates. We found a two-fold increase of mobility and current with tensile strain up to 0.7%, and a gauge factor (GF) up to ~200, which is the highest reported to date for sub-1 nm thin piezoresistive films. The improvements are attributed to changes in the band structure, including lower electron-phonon intervalley scattering and lower electron effective mass with tensile strain. These results achieve the largest mobility improvements of $MoS_2$ transistors with strain to date, pointing the way for performance enhancements in integrated 2D electronics, and for the use of this material in strain sensors on flexible substrates. For electronics on rigid substrates (i.e. integrated with silicon), strain could be induced with nitride or metal layers, or by growth on substrates with different thermal coefficients of expansion or lattice constants.[69-75] Some approaches would have the additional advantage of inducing biaxial strain in the 2D material, which is expected to have a larger effect on the electrical properties than uniaxial strain.[12,23,76-77]

## ASSOCIATED CONTENT

**Supporting Information.**

The following file is available free of charge on the ACS Publications website. Device fabrication and $MoS_2$ transfer process; Raman and photoluminescence (PL) spectroscopy of devices with strain; current-voltage characteristics of $MoS_2$ transistors with strain; *C-V* characteristics of Au-$Al_2O_3$-Au with strain; mobility ($\mu_{FE}$) and drain current ($I_D$) as a function of strain; and degradation of devices at high levels of strain.

## AUTHOR INFORMATION

**Corresponding Author**

*E-mail: epop@stanford.edu

**Present Addresses**

RWTH Aachen University (A.D.), Southern Methodist University (K.B.).

**Author Contributions**

The manuscript was written through contributions of all authors. All authors have given approval to the final version of the manuscript.

**Notes**

The authors declare no competing financial interest.

## ACKNOWLEDGMENTS

This work was performed in part at the Stanford Nanofabrication Facility (SNF) and the Stanford Nano Shared Facilities (SNSF) which receive funding from the National Science Foundation (NSF) as part of the NNCI award 1542152. This work was also supported by the NSF EFRI 2-DARE grant 1542883. I.M.D. acknowledges support from the NDSEG and ARCS Fellowships. A.D. has been in part supported by the Swiss National Science Foundation's Early Postdoc.Mobility fellowship (grant: P2EZP2_181619) and in part by Beijing Institute of Collaborative Innovation (BICI).

## REFERENCES

(1) Das, S.; Sebastian, A.; Pop, E.; McClellan, C. J.; Franklin, A. D.; Grasser, T.; Knobloch, T.; Illarionov, Y.; Penumatcha, A. V.; Appenzeller, J.; Chen, Z.; Zhu, W.; Asselberghs, I.; Li, L.-J.; Avci, U. E.; Bhat, N.; Anthopoulos, T. D.; Singh, R. Transistors based on two-dimensional materials for future integrated circuits. *Nature Electronics* **2021**, *4*, 786-799.

(2) Smithe, K. K. H.; English, C. D.; Suryavanshi, S. V.; Pop, E. Intrinsic electrical transport and performance projections of synthetic monolayer $MoS_2$ devices. *2D Materials* **2017**, *4*, 011009.

(3) Cun, H.; Macha, M.; Kim, H.; Liu, K.; Zhao, Y.; LaGrange, T.; Kis, A.; Radenovic, A. Wafer-scale MOCVD growth of monolayer $MoS_2$ on sapphire and $SiO_2$. *Nano Research* **2019**, *12*, 2646.

(4) Mirabelli, G.; McGeough, C.; Schmidt, M.; McCarthy, E. K.; Monaghan, S.; Povey, I. M.; McCarthy, M.; Gity, F.; Nagle, R.; Hughes, G.; Cafolla, A.; Hurley, P. K.; Duffy, R. Air sensitivity of $MoS_2$, $MoSe_2$, $MoTe_2$, $HfS_2$, and $HfSe_2$. *Journal of Applied Physics* **2016**, *120*, 125102.

(5) Das, S.; Chen, H.-Y.; Penumatcha, A. V.; Appenzeller, J. High Performance Multilayer $MoS_2$ Transistors with Scandium Contacts. *Nano Letters* **2013**, *13*, 100-105.

(6) English, C. D.; Shine, G.; Dorgan, V. E.; Saraswat, K. C.; Pop, E. Improved Contacts to $MoS_2$ Transistors by Ultra-High Vacuum Metal Deposition. *Nano Letters* **2016**, *16*, 3824-3830.

(7) Yang, L.; Majumdar, K.; Liu, H.; Du, Y.; Wu, H.; Hatzistergos, M.; Hung, P. Y.; Tieckelmann, R.; Tsai, W.; Hobbs, C.; Ye, P. D. Chloride Molecular Doping Technique on 2D Materials: $WS_2$ and $MoS_2$. *Nano Letters* **2014**, *14*, 6275-6280.

(8) Zhao, Y.; Xu, K.; Pan, F.; Zhou, C.; Zhou, F.; Chai, Y. Doping, Contact and Interface Engineering of Two-Dimensional Layered Transition Metal Dichalcogenides Transistors. *Advanced Functional Materials* **2017**, *27*, 1603484.

(9) McClellan, C. J.; Yalon, E.; Smithe, K. K. H.; Suryavanshi, S. V.; Pop, E. High Current Density in Monolayer $MoS_2$ Doped by $AlO_x$. *ACS Nano* **2021**, *15*, 1587-1596.

(10) Yu, Z.; Pan, Y.; Shen, Y.; Wang, Z.; Ong, Z.-Y.; Xu, T.; Xin, R.; Pan, L.; Wang, B.; Sun, L.; Wang, J.; Zhang, G.; Zhang, Y. W.; Shi, Y.; Wang, X. Towards intrinsic charge transport in monolayer molybdenum disulfide by defect and interface engineering. *Nature Communications* **2014,** *5,* 5290.

(11) Yu, Z.; Ong, Z.-Y.; Pan, Y.; Cui, Y.; Xin, R.; Shi, Y.; Wang, B.; Wu, Y.; Chen, T.; Zhang, Y.-W.; Zhang, G.; Wang, X. Realization of Room-Temperature Phonon-Limited Carrier Transport in Monolayer $MoS_2$ by Dielectric and Carrier Screening. *Advanced Materials* **2016,** *28,* 547-552.

(12) Hosseini, M.; Elahi, M.; Pourfath, M.; Esseni, D. Strain induced mobility modulation in single-layer $MoS_2$. *Journal of Physics D: Applied Physics* **2015,** *48,* 375104.

(13) Hosseini, M.; Elahi, M.; Pourfath, M.; Esseni, D. Strain-Induced Modulation of Electron Mobility in Single-Layer Transition Metal Dichalcogenides $MX_2$ (M = Mo, W; X = S, Se). *IEEE Transactions on Electron Devices* **2015,** *62,* 3192-3198.

(14) Harada, N.; Sato, S.; Yokoyama, N. Computational study on electrical properties of transition metal dichalcogenide field-effect transistors with strained channel. *Journal of Applied Physics* **2014,** *115,* 034505.

(15) Shi, H.; Pan, H.; Zhang, Y.-W.; Yakobson, B. I. Quasiparticle band structures and optical properties of strained monolayer $MoS_2$ and $WS_2$. *Physical Review B* **2013,** *87,* 155304.

(16) Takagi, S. i.; Hoyt, J. L.; Welser, J. J.; Gibbons, J. F. Comparative study of phonon-limited mobility of two-dimensional electrons in strained and unstrained Si metal–oxide–semiconductor field-effect transistors. *Journal of Applied Physics* **1996,** *80,* 1567-1577.

(17) Welser; Hoyt; Gibbons NMOS and PMOS transistors fabricated in strained silicon/relaxed silicon-germanium structures. *1992 International Technical Digest on Electron Devices Meeting*, 13-16 Dec. 1992, **1992.** 10.1109/IEDM.1992.307527.

(18) Thompson, S.; Anand, N.; Armstrong, M.; Auth, C.; Arcot, B.; Alavi, M.; Bai, P.; Bielefeld, J.; Bigwood, R.; Brandenburg, J.; Buehler, M.; Cea, S.; Chikarmane, V.; Choi, C.; Frankovic, R.; Ghani, T.; Glass, G.; Han, W.; Hoffmann, T.; Hussein, M.; Jacob, P.; Jain, A.; Jan, C.; Joshi, S.; Kenyon, C.; Klaus, J.; Klopcic, S.; Luce, J.; Ma, Z.; Mcintyre, B.; Mistry, K.; Murthy, A.; Nguyen, P.; Pearson, H.; Sandford, T.; Schweinfurth, R.; Shaheed, R.; Sivakumar, S.; Taylor, M.; Tufts, B.; Wallace, C.; Wang, P.; Weber, C.; Bohr, M. A 90 nm logic technology featuring 50 nm strained silicon channel transistors, 7 layers of Cu interconnects, low κ ILD, and 1 $\mu m^2$ SRAM cell. *International Electron Devices Meeting (IEDM) Technical Digest*, 8-11 Dec. 2002, **2002.**

(19) Mistry, K.; Armstrong, M.; Auth, C.; Cea, S.; Coan, T.; Ghani, T.; Hoffmann, T.; Murthy, A.; Sandford, J.; Shaheed, R.; Zawadzki, K.; Zhang, K.; Thompson, S.; Bohr, M. Delaying forever: Uniaxial strained silicon transistors in a 90nm CMOS technology. *Symposium on VLSI Technology*, **2004.**

(20) Thompson, S. E.; Armstrong, M.; Auth, C.; Cea, S.; Chau, R.; Glass, G.; Hoffman, T.; Klaus, J.; Zhiyong, M.; Mcintyre, B.; Murthy, A.; Obradovic, B.; Shifren, L.; Sivakumar, S.; Tyagi, S.; Ghani, T.; Mistry, K.; Bohr, M.; El-Mansy, Y. A logic nanotechnology featuring strained-silicon. *IEEE Electron Device Letters* **2004,** *25,* 191-193.

(21) Mohta, N.; Thompson, S. E. Mobility enhancement: The next vector to extend Moore's law. *IEEE Circuits Devices Mag.* **2005,** *21,* 18-23.

(22) Rice, C.; Young, R. J.; Zan, R.; Bangert, U.; Wolverson, D.; Georgiou, T.; Jalil, R.; Novoselov, K. S. Raman-scattering measurements and first-principles calculations of strain-induced phonon shifts in monolayer $MoS_2$. *Physical Review B* **2013,** *87,* 081307.

(23) Chang, C.-H.; Fan, X.; Lin, S.-H.; Kuo, J.-L. Orbital analysis of electronic structure and phonon dispersion in $MoS_2$, $MoSe_2$, $WS_2$, and $WSe_2$ monolayers under strain. *Physical Review B* **2013,** *88,* 195420.

(24) He, K.; Poole, C.; Mak, K. F.; Shan, J. Experimental Demonstration of Continuous Electronic Structure Tuning via Strain in Atomically Thin $MoS_2$. *Nano Letters* **2013**, *13*, 2931-2936.

(25) Aslan, O. B.; Datye, I. M.; Mleczko, M. J.; Sze Cheung, K.; Krylyuk, S.; Bruma, A.; Kalish, I.; Davydov, A. V.; Pop, E.; Heinz, T. F. Probing the Optical Properties and Strain-Tuning of Ultrathin $Mo_{1-x}W_xTe_2$. *Nano Letters* **2018**, *18*, 2485-2491.

(26) Conley, H. J.; Wang, B.; Ziegler, J. I.; Haglund, R. F.; Pantelides, S. T.; Bolotin, K. I. Bandgap Engineering of Strained Monolayer and Bilayer $MoS_2$. *Nano Letters* **2013**, *13*, 3626-3630.

(27) Vaziri, S.; Yalon, E.; Munoz Rojo, M.; Suryavanshi, S. V.; Zhang, H.; McClellan, C. J.; Bailey, C. S.; Smithe, K. K. H.; Gabourie, A. J.; Chen, V.; Deshmukh, S.; Bendersky, L.; Davydov, A. V.; Pop, E. Ultrahigh thermal isolation across heterogeneously layered two-dimensional materials. *Sci Adv* **2019**, *5*, eaax1325.

(28) Bertolazzi, S.; Brivio, J.; Kis, A. Stretching and Breaking of Ultrathin $MoS_2$. *ACS Nano* **2011**, *5*, 9703-9709.

(29) Mohiuddin, T. M. G.; Lombardo, A.; Nair, R. R.; Bonetti, A.; Savini, G.; Jalil, R.; Bonini, N.; Basko, D. M.; Galiotis, C.; Marzari, N.; Novoselov, K. S.; Geim, A. K.; Ferrari, A. C. Uniaxial strain in graphene by Raman spectroscopy: G peak splitting, Gruneisen parameters, and sample orientation. *Physical Review B* **2009**, *79*, 205433.

(30) Schauble, K.; Zakhidov, D.; Yalon, E.; Deshmukh, S.; Grady, R. W.; Cooley, K. A.; McClellan, C. J.; Vaziri, S.; Passarello, D.; Mohney, S. E.; Toney, M. F.; Sood, A. K.; Salleo, A.; Pop, E. Uncovering the Effects of Metal Contacts on Monolayer $MoS_2$. *ACS Nano* **2020**, *14*, 14798–14808.

(31) Smithe, K. K. H.; Suryavanshi, S. V.; Muñoz Rojo, M.; Tedjarati, A. D.; Pop, E. Low Variability in Synthetic Monolayer $MoS_2$ Devices. *ACS Nano* **2017**, *11*, 8456-8463.

(32) Christopher, J. W.; Vutukuru, M.; Lloyd, D.; Bunch, J. S.; Goldberg, B. B.; Bishop, D. J.; Swan, A. K. Monolayer $MoS_2$ Strained to 1.3% With a Microelectromechanical System. *Journal of Microelectromechanical Systems* **2019**, *28*, 254-263.

(33) John, A. P.; Thenapparambil, A.; Thalakulam, M. Strain-engineering the Schottky barrier and electrical transport on $MoS_2$. *Nanotechnology* **2020**, *31*, 275703.

(34) Zhu, C. R.; Wang, G.; Liu, B. L.; Marie, X.; Qiao, X. F.; Zhang, X.; Wu, X. X.; Fan, H.; Tan, P. H.; Amand, T.; Urbaszek, B. Strain tuning of optical emission energy and polarization in monolayer and bilayer $MoS_2$. *Physical Review B* **2013**, *88*, 121301.

(35) Beal, A. R.; Knights, J. C.; Liang, W. Y. Transmission spectra of some transition metal dichalcogenides. II. Group VIA: trigonal prismatic coordination. *Journal of Physics C: Solid State Physics* **1972**, *5*, 3540-3551.

(36) Wilson, J. A.; Yoffe, A. D. The transition metal dichalcogenides discussion and interpretation of the observed optical, electrical and structural properties. *Advances in Physics* **1969**, *18*, 193-335.

(37) Ji-Song, L.; Thompson, S. E.; Fossum, J. G. Comparison of threshold-voltage shifts for uniaxial and biaxial tensile-stressed n-MOSFETs. *IEEE Electron Device Letters* **2004**, *25*, 731-733.

(38) Dobrescu, L.; Petrov, M.; Dobrescu, D.; Ravariu, C. Threshold voltage extraction methods for MOS transistors. *2000 International Semiconductor Conference*, 10-14 Oct. 2000, **2000**. 10.1109/SMICND.2000.890257.

(39)  Chang, H.-Y.; Yang, S.; Lee, J.; Tao, L.; Hwang, W.-S.; Jena, D.; Lu, N.; Akinwande, D. High-Performance, Highly Bendable MoS2 Transistors with High-κ Dielectrics for Flexible Low-Power Systems. *ACS Nano* **2013**, *7*, 5446-5452.

(40)  Münzenrieder, N.; Petti, L.; Zysset, C.; Görk, D.; Büthe, L.; Salvatore, G. A.; Tröster, G. Investigation of gate material ductility enables flexible a-IGZO TFTs bendable to a radius of 1.7 mm. *2013 Proceedings of the European Solid-State Device Research Conference (ESSDERC)*, 16-20 Sept. 2013, **2013**. 10.1109/ESSDERC.2013.6818893.

(41)  Ferry, D. K. Electron transport in some transition metal di-chalcogenides: MoS2 and WS2. *Semiconductor Science and Technology* **2017**, *32*, 085003.

(42)  Herring, C. Character tables for two space groups. *Journal of the Franklin Institute* **1942**, *233*, 525-543.

(43)  Ribeiro-Soares, J.; Almeida, R. M.; Barros, E. B.; Araujo, P. T.; Dresselhaus, M. S.; Cançado, L. G.; Jorio, A. Group theory analysis of phonons in two-dimensional transition metal dichalcogenides. *Physical Review B* **2014**, *90*, 115438.

(44)  Hill, H. M.; Rigosi, A. F.; Rim, K. T.; Flynn, G. W.; Heinz, T. F. Band Alignment in MoS2/WS2 Transition Metal Dichalcogenide Heterostructures Probed by Scanning Tunneling Microscopy and Spectroscopy. *Nano Letters* **2016**, *16*, 4831-4837.

(45)  Gaddemane, G.; Gopalan, S.; Van de Put, M. L.; Fischetti, M. V. Limitations of ab initio methods to predict the electronic-transport properties of two-dimensional semiconductors: the computational example of 2H-phase transition metal dichalcogenides. *Journal of Computational Electronics* **2021**, *20*, 49-59.

(46)  Jin, Z.; Li, X.; Mullen, J. T.; Kim, K. W. Intrinsic transport properties of electrons and holes in monolayer transition-metal dichalcogenides. *Physical Review B* **2014**, *90*, 045422.

(47)  Aslan, O. B.; Deng, M.; Heinz, T. F. Strain tuning of excitons in monolayer WSe2. *Physical Review B* **2018**, *98*, 115308.

(48)  Yue, Q.; Kang, J.; Shao, Z.; Zhang, X.; Chang, S.; Wang, G.; Qin, S.; Li, J. Mechanical and electronic properties of monolayer MoS2 under elastic strain. *Physics Letters A* **2012**, *376*, 1166-1170.

(49)  Yun, W. S.; Han, S. W.; Hong, S. C.; Kim, I. G.; Lee, J. D. Thickness and strain effects on electronic structures of transition metal dichalcogenides: 2H-MX2 semiconductors (M = Mo, W; X = S, Se, Te). *Physical Review B* **2012**, *85*, 033305.

(50)  Quereda, J.; Palacios, J. J.; Agräit, N.; Castellanos-Gomez, A.; Rubio-Bollinger, G. Strain engineering of Schottky barriers in single- and few-layer MoS2 vertical devices. *2D Materials* **2017**, *4*, 021006.

(51)  Shen, T.; Penumatcha, A. V.; Appenzeller, J. Strain Engineering for Transition Metal Dichalcogenides Based Field Effect Transistors. *ACS Nano* **2016**, *10*, 4712-4718.

(52)  Kumar, A.; Schauble, K.; Neilson, K. M.; Tang, A.; Ramesh, P.; Wong, H. S. P.; Pop, E.; Saraswat, K. Sub-200 Ω•um Alloyed Contacts to Synthetic Monolayer MoS2. *2021 IEEE International Electron Devices Meeting (IEDM)*, 11-16 Dec. 2021, **2021**. 10.1109/IEDM19574.2021.9720609.

(53)  Zhu, W.; Low, T.; Lee, Y.-H.; Wang, H.; Farmer, D. B.; Kong, J.; Xia, F.; Avouris, P. Electronic transport and device prospects of monolayer molybdenum disulphide grown by chemical vapour deposition. *Nature Communications* **2014**, *5*, 3087.

(54) Hong, H.; Cheng, Y.; Wu, C.; Huang, C.; Liu, C.; Yu, W.; Zhou, X.; Ma, C.; Wang, J.; Zhang, Z.; Zhao, Y.; Xiong, J.; Liu, K. Modulation of carrier lifetime in MoS2 monolayer by uniaxial strain. *Chinese Physics B* **2020**, *29*, 077201.

(55) Beckwith, T. G.; Marangoni, R. D.; Lienhard, J. H., *Mechanical Measurements*. International Edition ed.; Pearson Education, Limited: 2008.

(56) Li, C.; Hesketh, P. J.; Maclay, G. J. Thin gold film strain gauges. *Journal of Vacuum Science & Technology A* **1994**, *12*, 813-819.

(57) French, P. J.; Evans, A. G. R. Piezoresistance in polysilicon and its applications to strain gauges. *Solid-State Electronics* **1989**, *32*, 1-10.

(58) Kim, Y.; Kim, Y.; Lee, C.; Kwon, S. Thin Polysilicon Gauge for Strain Measurement of Structural Elements. *IEEE Sensors Journal* **2010**, *10*, 1320-1327.

(59) Beeby, S., *MEMS Mechanical Sensors*. Artech House: 2004.

(60) Smith, C. S. Piezoresistance Effect in Germanium and Silicon. *Physical Review* **1954**, *94*, 42-49.

(61) Manzeli, S.; Allain, A.; Ghadimi, A.; Kis, A. Piezoresistivity and Strain-induced Band Gap Tuning in Atomically Thin MoS2. *Nano Letters* **2015**, *15*, 5330-5335.

(62) Park, M.; Park, Y. J.; Chen, X.; Park, Y.-K.; Kim, M.-S.; Ahn, J.-H. MoS2-Based Tactile Sensor for Electronic Skin Applications. *Advanced Materials* **2016**, *28*, 2556-2562.

(63) Zhu, M.; Li, J.; Inomata, N.; Toda, M.; Ono, T. Vanadium-doped molybdenum disulfide film-based strain sensors with high gauge factor. *Applied Physics Express* **2019**, *12*, 015003.

(64) Zhu, M.; Sakamoto, K.; Li, J.; Inomata, N.; Toda, M.; Ono, T. Piezoresistive strain sensor based on monolayer molybdenum disulfide continuous film deposited by chemical vapor deposition. *Journal of Micromechanics and Microengineering* **2019**, *29*, 055002.

(65) Chen, L.; Liang, D.; Yu, Z.; Li, S.; Feng, X.; Li, B.; Li, Y.; Zhang, Y.; Ang, K. Ultrasensitive Flexible Strain Sensor based on Two-Dimensional InSe for Human Motion Surveillance. *2019 IEEE International Electron Devices Meeting (IEDM)*, 7-11 Dec. 2019, **2019.** 10.1109/IEDM19573.2019.8993476.

(66) Wagner, S.; Yim, C.; McEvoy, N.; Kataria, S.; Yokaribas, V.; Kuc, A.; Pindl, S.; Fritzen, C.-P.; Heine, T.; Duesberg, G. S.; Lemme, M. C. Highly Sensitive Electromechanical Piezoresistive Pressure Sensors Based on Large-Area Layered PtSe2 Films. *Nano Letters* **2018**, *18*, 3738-3745.

(67) Zhang, Z.; Li, L.; Horng, J.; Wang, N. Z.; Yang, F.; Yu, Y.; Zhang, Y.; Chen, G.; Watanabe, K.; Taniguchi, T.; Chen, X. H.; Wang, F.; Zhang, Y. Strain-Modulated Bandgap and Piezo-Resistive Effect in Black Phosphorus Field-Effect Transistors. *Nano Letters* **2017**, *17*, 6097-6103.

(68) Won, S. M.; Kim, H.; Lu, N.; Kim, D.; Solar, C. D.; Duenas, T.; Ameen, A.; Rogers, J. A. Piezoresistive Strain Sensors and Multiplexed Arrays Using Assemblies of Single-Crystalline Silicon Nanoribbons on Plastic Substrates. *IEEE Transactions on Electron Devices* **2011**, *58*, 4074-4078.

(69) Wan, W.; Chen, L.; Zhan, L.; Zhu, Z.; Zhou, Y.; Shih, T.; Guo, S.; Kang, J.; Huang, H.; Cai, W. Syntheses and bandgap alterations of MoS2 induced by stresses in graphene-platinum substrates. *Carbon* **2018**, *131*, 26-30.

(70) Ahn, G. H.; Amani, M.; Rasool, H.; Lien, D.-H.; Mastandrea, J. P.; Ager Iii, J. W.; Dubey, M.; Chrzan, D. C.; Minor, A. M.; Javey, A. Strain-engineered growth of two-dimensional materials. *Nature Communications* **2017**, *8*, 608.

(71)  Chae, W. H.; Cain, J. D.; Hanson, E. D.; Murthy, A. A.; Dravid, V. P. Substrate-induced strain and charge doping in CVD-grown monolayer $MoS_2$. *Applied Physics Letters* **2017,** *111*, 143106.

(72)  Chai, Y.; Su, S.; Yan, D.; Ozkan, M.; Lake, R.; Ozkan, C. S. Strain Gated Bilayer Molybdenum Disulfide Field Effect Transistor with Edge Contacts. *Scientific Reports* **2017,** *7*, 41593.

(73)  Chen, P.; Xu, W.; Gao, Y.; Warner, J. H.; Castell, M. R. Epitaxial Growth of Monolayer $MoS_2$ on $SrTiO_3$ Single Crystal Substrates for Applications in Nanoelectronics. *ACS Applied Nano Materials* **2018,** *1*, 6976-6988.

(74)  Peña, T.; Chowdhury, S. A.; Azizimanesh, A.; Sewaket, A.; Askari, H.; Wu, S. M. Strain engineering 2D $MoS_2$ with thin film stress capping layers. *2D Materials* **2021,** *8*, 045001.

(75)  Chen, Y.; Deng, W.; Chen, X.; Wu, Y.; Shi, J.; Zheng, J.; Chu, F.; Liu, B.; An, B.; You, C.; Jiao, L.; Liu, X.; Zhang, Y. Carrier mobility tuning of $MoS2$ by strain engineering in CVD growth process. *Nano Research* **2021,** *14*, 2314-2320.

(76)  Ghorbani-Asl, M.; Borini, S.; Kuc, A.; Heine, T. Strain-dependent modulation of conductivity in single-layer transition-metal dichalcogenides. *Physical Review B* **2013,** *87*, 235434.

(77)  Johari, P.; Shenoy, V. B. Tuning the Electronic Properties of Semiconducting Transition Metal Dichalcogenides by Applying Mechanical Strains. *ACS Nano* **2012,** *6*, 5449-5456.

**Table of Contents Figure**

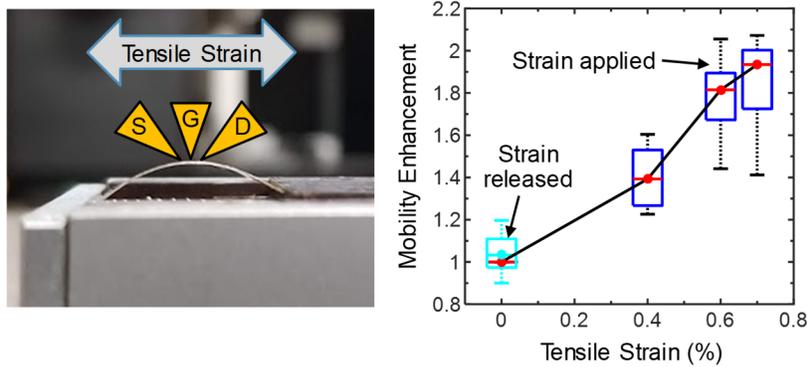

**Supporting Information**

# Strain-Enhanced Mobility of Monolayer MoS₂

Isha M. Datye,[1] Alwin Daus,[1,2] Ryan W. Grady,[1] Kevin Brenner,[1,3] Sam Vaziri,[1] and Eric Pop[1,4,5,*]

[1]*Department of Electrical Engineering, Stanford University, Stanford, CA 94305, USA*
[2]*Present address: Chair for Electronic Devices, RWTH Aachen University, Aachen, 52074, Germany*
[3]*Present address: Department of Electrical and Computer Engineering, Southern Methodist University, Dallas, TX 75275, USA*
[4]*Department of Materials Science & Engineering, Stanford University, Stanford, CA 94305, USA*
[5]*Precourt Institute for Energy, Stanford University, Stanford, CA 94305, USA*

*Corresponding Author: epop@stanford.edu

## 1. Device fabrication and MoS₂ transfer process

### 1.1 Fabrication of monolayer (1L) MoS₂ transistors

The steps for fabricating our MoS₂ transistors on polyethylene naphthalate (PEN) substrates are described below. We include a schematic for each step of the process in Figure S1.

First, we pattern Cu (28 nm)/Au (5 nm) back-gates (BG) by optical lithography directly on the PEN (125 μm thick), and deposit them using electron-beam (e-beam) evaporation. We use Cu in contact with PEN because of its higher ductility than Au,[1] and we use a thin Au layer to encapsulate the Cu and limit its oxidation. We then deposit ~20 nm aluminum oxide ($Al_2O_3$) using atomic layer deposition (ALD) in a Savannah S200 from Cambridge Nanotech, with trimethyl aluminum (TMA) and $H_2O$ as precursors for 200 cycles. We perform ALD at 130°C, which is within the thermal limitations of our PEN substrates.

The MoS₂ is grown by chemical vapor deposition (CVD) on separate SiO₂/Si substrates[2] (chips of approx. 1.5×2 cm) and transferred onto the $Al_2O_3$, as detailed in Section 1.2. Next, we pattern and e-beam evaporate Au (55 nm) to form the source and drain contacts to the transistors. The channel length of our transistors ranges from 2 to 15 μm, which is within the limits of optical lithography on PEN substrates. We also define and evaporate Ti (3 nm)/Cu (30 nm)/Au (5 nm) pads connected to the Au contacts for device probing. We note that before metal deposition of the pads, we etch the MoS₂ underneath this region to improve the adhesion of the metal to the PEN substrate. The MoS₂ is etched using O₂ plasma with 100 sccm O₂ at 150 mTorr at a power of 50 W for 45 seconds. Finally, the MoS₂ channel is patterned and etched into a rectangle to define the width of the transistor channel using the same etch conditions. We note that the fabrication processes outlined here do not damage the MoS₂ film or leave photoresist residue.[3]

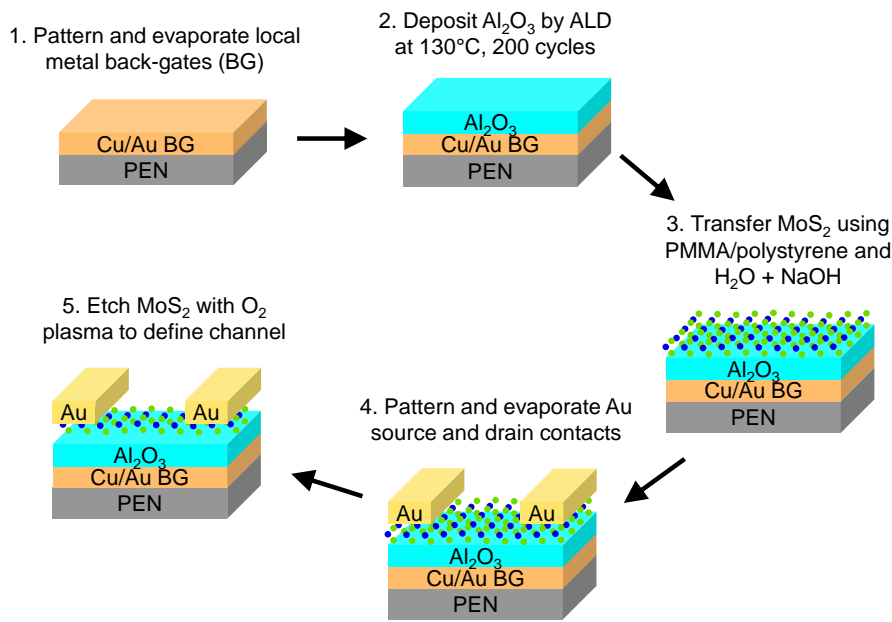

**Figure S1.** Summary of steps used to fabricate our MoS$_2$ transistors on flexible PEN substrates.

## 1.2 MoS$_2$ transfer process

We describe the steps used to transfer MoS$_2$ onto PEN substrates as follows.[4]

1. Place thin strips of tape (Nitto Denko RevAlpha series thermal release tape or Kapton) on all four edges of the chip with MoS$_2$ grown by CVD on SiO$_2$/Si.

2. Spin-coat 2% PMMA at 2500 rpm for 1 minute, and then bake for 45 seconds at 135°C.

3. Spin-coat polystyrene (PS) dissolved in toluene (PS/toluene 3 g/20 mL) at 2000 rpm for 1 minute, and then bake for 5 minutes at 85°C.

4. Remove the tape very carefully from edges of the chip.

5. Place MoS$_2$ sample with polymer support layers in a beaker of water. Agitate sample in water to cause delamination of MoS$_2$/polymer stack from growth substrate. Use tweezers to poke edges of the polymer stack to initiate delamination from substrate.

6. If MoS$_2$/polymer stack does not delaminate in water, place sample in a beaker of 1M NaOH for a few minutes to etch SiO$_2$ and promote delamination from the substrate. After the MoS$_2$/polymer stack begins to delaminate from the SiO$_2$/Si substrate, return the chip to a fresh beaker of water. Return to step 5 until the entire film has delaminated from the substrate.

7. After the MoS$_2$/polymer film has delaminated from the growth substrate, it will float on the water. Use the PEN to pick up the floating sample.

8. Carefully blow a N$_2$ gun perpendicular to the sample to remove water trapped between the MoS$_2$ and PEN.

9. Place the sample on a hot plate at 55°C for 5 minutes, then increase the temperature to 85°C and heat the sample for 5 minutes, and finally increase the temperature to 135°C. This gradual increase in temperature minimizes damage (e.g. bubbles and holes) to the MoS$_2$ film after the transfer. When the temperature reaches 135°C, remove the sample from the hot plate.

10. Place the obtained polymer/MoS$_2$/PEN stack in toluene for 12 hours to dissolve the PMMA and PS polymers. Rinse the sample in acetone and IPA after removing it from toluene.

## 2. Raman and photoluminescence (PL) spectroscopy of devices with strain

We use Raman and photoluminescence (PL) spectroscopy to confirm that the MoS$_2$ is under tensile strain, in addition to the visual inspections shown in Figure 2b of the main text.

Figure S2a-b below illustrates the E' (in-plane) and A$_1$' (out-of-plane) Raman peak positions for 10 devices, respectively. The data are presented as box plots for 0%, 0.7%, and back to 0% applied tensile strain. For each transistor, we averaged the Raman peak positions over ~5 spots across the channel. The average E' peak positions across all 10 devices after device fabrication, with 0.7% strain, and after the strain is released are ~384.6 ± 0.1 cm$^{-1}$, ~382.8 ± 0.4 cm$^{-1}$, and ~384.6 ± 0.2 cm$^{-1}$, respectively. This marks a 2.5 ± 0.5 cm$^{-1}$/% strain shifting rate of the E' peak.

As expected, the shift of the A$_1$' Raman peak (at ~403 cm$^{-1}$) is lower (1.0 ± 0.3 cm$^{-1}$/% strain) than that of the E' peak. This shift may be partially due to altered substrate interactions rather than purely a direct strain effect on the out-of-plane vibrations.[5]

We note that some studies report peak splitting of the degenerate E' Raman mode with tensile strain due to breaking the symmetry of the crystal.[6-7] Our MoS$_2$ E' peak did not exhibit such behavior, likely because the MoS$_2$ did not experience large enough strains.

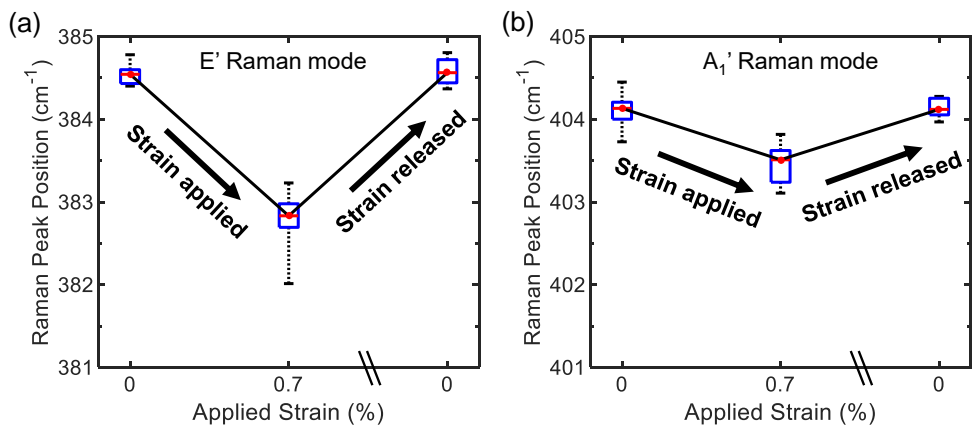

**Figure S2.** Box plots showing (a) E' and (b) A$_1$' Raman peak positions. The figures include data averaged from ~5 spots across the channels of all devices and show Raman peak positions before applying strain, with 0.7% strain, and after the strain returns to 0%.

We also perform PL spectroscopy on our samples to study the change in the optical band gap as a function of strain, with the A and B excitons schematically illustrated in Figure S3a. Figure S3b shows the position of the A exciton peak at 0%, 0.4%, 0.6%, and back to 0% strain. Each box plot represents data from 7 transistors, and for each transistor we averaged the peak position over 3 spots in the channel region. The average A exciton peak positions at 0% strain, 0.4% strain, 0.6% strain, and back to 0% strain are 1.813 ± 0.004 eV, 1.793 ± 0.005 eV, 1.773 ± 0.005 eV, and 1.814 ± 0.005 eV, respectively. We also point out that the PL peak intensity decreases slightly with increasing tensile strain (see Figures 2d-e in the main text), due to a narrowing of the indirect optical gap of monolayer MoS$_2$ between the Γ (valence) and K (conduction) points.[6,8]

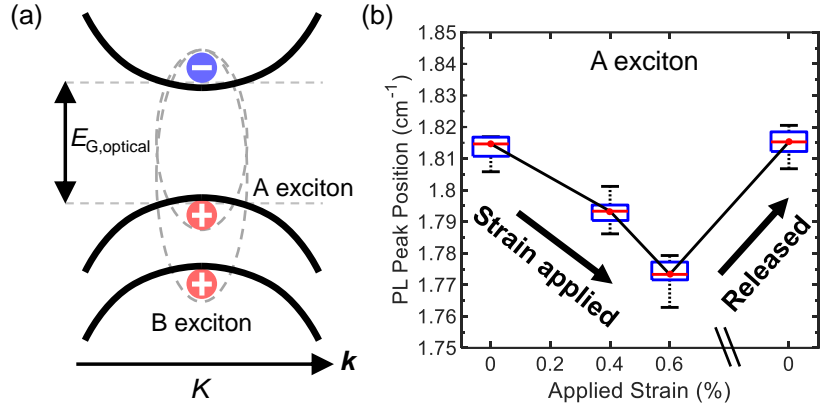

**Figure S3.** (a) Schematic band diagram of monolayer MoS₂ showing the optical band gap ($E_G$,optical) and the A and B excitons at the K point, without strain applied.[9-10] (b) Box plots showing the A exciton peak position from photoluminescence measurements as a function of strain, including after the strain is released. Each box plot includes data from 3 spots across the channels of 7 devices.

We note that CVD-grown MoS₂ typically attains some built-in tensile strain during the growth process, due to the larger thermal coefficient of expansion (TCE) of MoS₂ compared to the underlying SiO₂.[11] Transferring the MoS₂ to another substrate should release this built-in strain, so we expect that the MoS₂ is either unstrained or slightly compressively strained upon transfer to the PEN. However, the Raman and PL peak positions of unstrained MoS₂ are difficult to determine because these peaks are affected by many factors in addition to strain, such as doping and substrate interactions.[5,11-13] Our MoS₂ is transferred onto Al₂O₃, which is known to cause peak shifts in the Raman and PL spectra of MoS₂.[14] Therefore, we are unable to conclude whether our devices initially have any built-in tensile or compressive strain from the transfer and fabrication processes.

## 3. Current-voltage characteristics of MoS₂ transistors with strain

We perform electrical measurements with a Keithley 4200-SCS using Keithley Interactive Test Environment (KITE) software. Figure S4 shows the same $I_D$-$V_{GS}$ curves from Figure 3a of the main text but here includes the reverse sweep (dashed lines) in addition to the forward sweep (solid lines). The difference between the forward and reverse voltage sweep measurements points to hysteresis in the device, likely arising from electrostatic screening by H₂O and O₂ adsorbates trapped at the MoS₂-Al₂O₃ interface during the transfer process.[15] The gate current ($I_G$) at different strain levels is depicted by the dotted lines and does not exceed 1.2 nA during any of the measurements, indicating that the leakage current through the gate dielectric remains low during all strain-dependent measurements.

The $I_D$-$V_{DS}$ measurements shown in Figure 3b of the main text include forward and backward voltage sweeps with no observable hysteresis, which also suggests that the traps are located at the MoS₂-Al₂O₃ interface or within the Al₂O₃, and are mainly affected by sweeping the gate voltage.

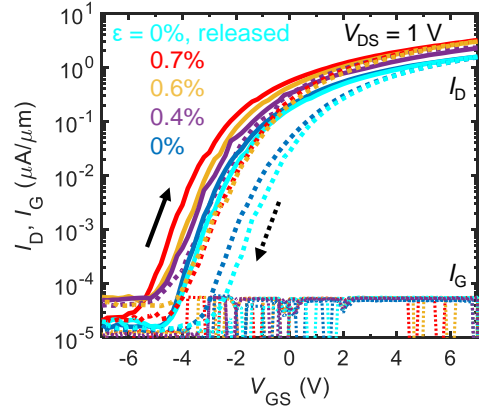

**Figure S4.** Measured $I_D$-$V_{GS}$ curves showing forward (solid) and backward (dashed lines) sweeps at different strain levels, for the same device shown in Figure 3a of the main text, with $W = 20$ μm, $L = 8$ μm, and $V_{DS} = 1$ V. The gate current ($I_G$) at different levels of strain is depicted by the dotted lines and does not exceed 1.2 nA for all measurements.

## 4. *C-V* characteristics of Au-Al₂O₃-Au capacitors with strain

We characterize the capacitance of the Al₂O₃ gate oxide using metal-oxide-metal structures (see Figure S5a for a top-view schematic) to enable more accurate estimation of field-effect mobility ($\mu_{FE}$). We verify that the capacitance does not change with strain, as shown in Figure S5b. We measure an oxide capacitance $C_{ox} \approx 312$ nF/cm², which we use in our calculation of $\mu_{FE}$. With ellipsometry, we estimate an Al₂O₃ thickness $t_{ox} \sim 20$ nm, which gives a relative dielectric constant $\epsilon_{ox} \sim 7$ for our Al₂O₃, based on the equation $C_{ox} = \epsilon_{ox}\epsilon_o/t_{ox}$, where $C_{ox}$ is normalized by the area of the Au electrode overlap and $\epsilon_o = 8.85 \times 10^{-14}$ F/cm. The equivalent oxide thickness ($EOT = \epsilon_{SiO2}t_{ox}/\epsilon_{ox}$) of our Al₂O₃ film is ~11 nm. The alternating current (AC) frequency and voltage bias during the measurements are 20 kHz and 30 mV, respectively.

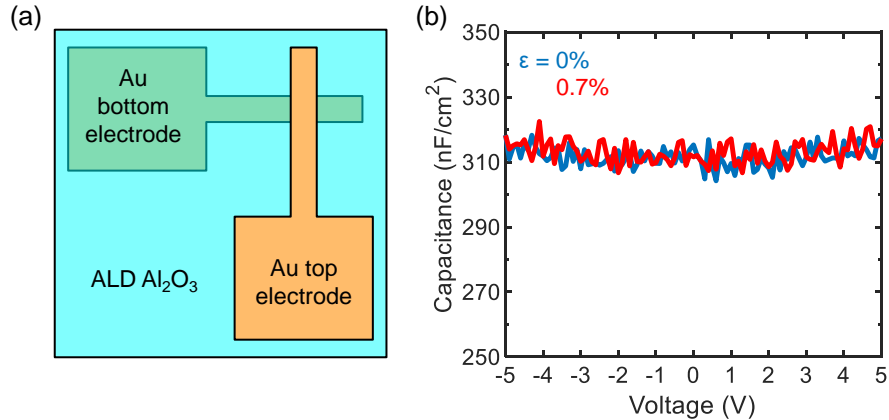

**Figure S5.** (a) Schematic of Au-Al₂O₃-Au structure for capacitance-voltage measurements. (b) Capacitance-voltage measurements of an Au-Al₂O₃-Au test structure with and without 0.7% tensile strain. The capacitance is normalized by the area of the top and bottom Au electrode overlap.

## 5. Mobility ($\mu_{FE}$) and drain current ($I_D$) as a function of tensile strain

Figure S6a-b shows the magnitudes of $\mu_{FE}$ and $I_D$ as a function of strain, with each color corresponding to a different device. The data for 8 devices shown here correspond to the same data

represented in box plots in Figure 3c-d of the main text. All values of $\mu_{FE}$ are extracted at the same carrier density $n \sim 1.1 \times 10^{13}$ cm$^{-2}$ and $V_{DS} = 1$ V, and all values of $I_D$ are extracted at $V_{GS} = 7$ V and $V_{DS} = 1$ V. The data points on the right side of each plot show $\mu_{FE}$ and $I_D$ after the strain is released back to 0%. We note that the unstrained mobilities are lower than in some other MoS$_2$ studies,[2,16] which we attribute to growth variations and partial degradation of the MoS$_2$ after transfer to the flexible substrates. We also perform strain-dependent electrical measurements on a different set of monolayer MoS$_2$ transistors with higher initial mobilities, which is discussed in Section 6.

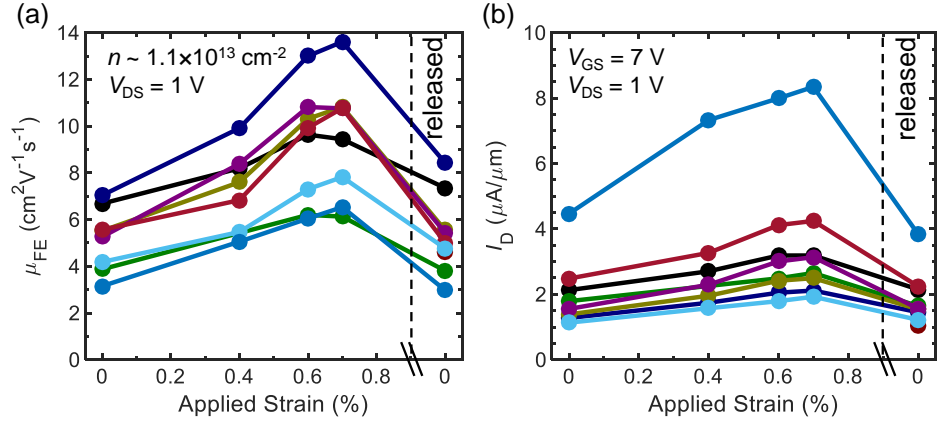

**Figure S6.** (a) Mobility ($\mu_{FE}$) and (b) drain current ($I_D$) for 8 different devices at 0%, 0.4%, 0.6%, 0.7%, and back to 0% strain. Channel lengths of these devices are $L = 2$ to 15 μm, also see Figure S7.

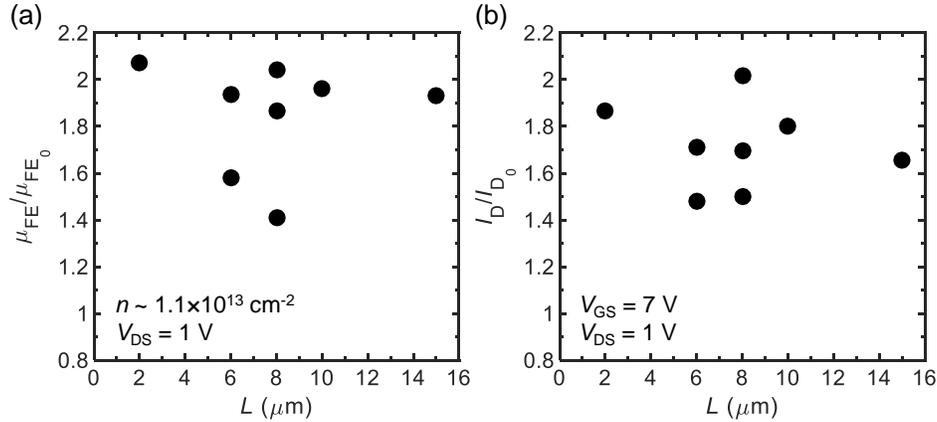

**Figure S7.** (a) Normalized $\mu_{FE}$ and (b) normalized $I_D$ at 0.7% tensile strain as a function of channel length. The mobility and current are normalized to their unstrained values (subscript "0").

We plot normalized $\mu_{FE}$ and $I_D$ at 0.7% strain as a function of $L = 2$ to 15 μm (see Figure S7a-b). There is no noticeable trend of mobility or current improvement with strain for different channel lengths, indicating that contact effects do not play a significant role across this range of channel lengths. Some theoretical studies have also predicted a directional (armchair vs. zigzag) dependence of MoS$_2$ mobility on strain.[17] However, the theoretical predictions are virtually indistinguishable at the lower strains applied in our work (<1%) and they are below our measurement sensitivity. In other words, we cannot conclude whether the variability observed in Figure S7 is due to applying strain along different crystallographic directions or (more likely) due to device and fabrication non-uniformity.

Figure S8a-b shows absolute and normalized $\mu_{FE}$ as a function of strain at a lower carrier density ($n \sim 4.8 \times 10^{12}$ cm$^{-2}$). The absolute and normalized $\mu_{FE}$ at lower $n$ are slightly lower than those at higher $n$ (see Figure S6a and Figure 3c in the main text), but the overall trends are similar.

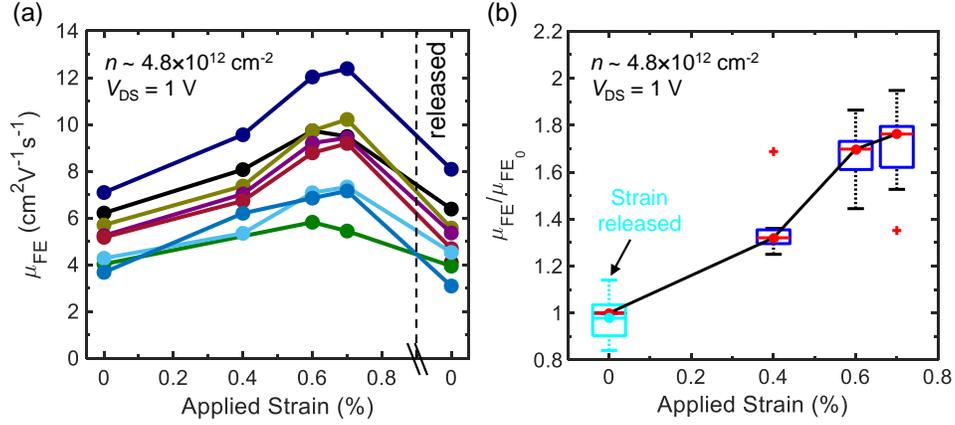

**Figure S8.** (a) Mobility ($\mu_{FE}$) and (b) normalized mobility ($\mu_{FE}/\mu_{FE,0}$) for the same 8 devices as in Figure S6-S7, at 0%, 0.4%, 0.6%, 0.7%, and back to 0% tensile strain at a lower carrier density $n \sim 4.8 \times 10^{12}$ cm$^{-2}$. The $\mu_{FE}$ in (b) is normalized to the initial (unstrained) values, with the box plots showing the median across devices (red circles), first and third quartiles (blue box), and maximum and minimum (top and bottom lines, respectively). The red "+" symbols represent outliers in the data. The cyan box plot corresponds to the measurement after strain is released.

## 6. Degradation of devices at high levels of strain

At strain greater than 0.7%, we observe a degradation of mobility and current in all devices. Figure S9a depicts this decrease in mobility at 0.8% strain during the 1st measurement of a second set of devices. When the strain returns to 0%, the mobility decreases further. However, the 2nd measurement (Figure S9b) shows an increase in mobility when strain is again applied to the devices, though it is lower than the improvement in mobility observed in Figure S6. These results demonstrate that there is a "break-in" of the device when the strain exceeds 0.7%, resulting in a degradation in performance.

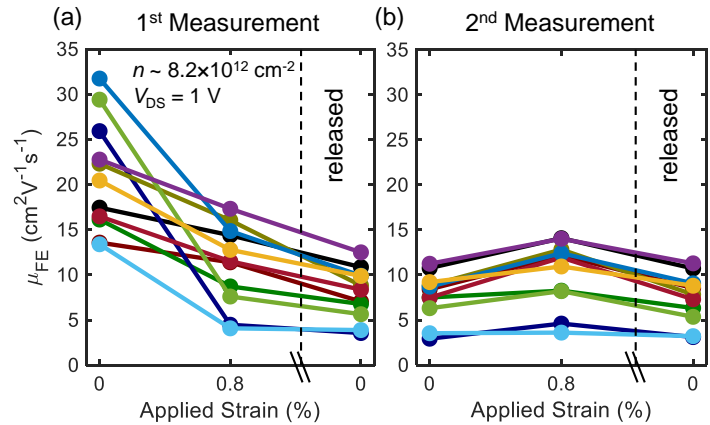

**Figure S9.** Field-effect mobility $\mu_{FE}$ as a function of larger tensile strain for devices on a different sample. Each color corresponds to a different device. (a) First measurement of devices at 0%, 0.8%, and back to 0% strain, showing a degradation in mobility. (b) Second measurement after device "break-in" at 0%, 0.8%, and back to 0% strain, this time showing an improvement in mobility at 0.8% strain, but lower mobility overall.

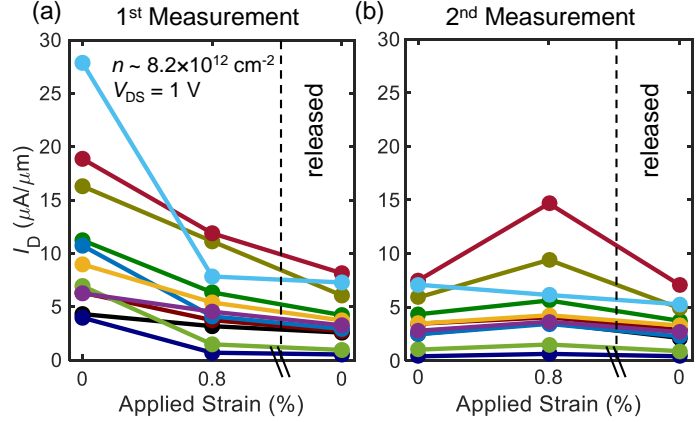

**Figure S10.** Current $I_D$ as a function of larger tensile strain for the same devices shown in Figure S9. Each color corresponds to a different device. (a) First measurement of devices at 0%, 0.8%, and back to 0% strain, showing a degradation in current. (b) Second measurement after device "break-in" at 0%, 0.8%, and back to 0% strain, this time showing an improvement in current at 0.8% strain.

Figure S10a-b shows the same trends of current $I_D$ at higher strain, i.e. an initial decrease in current with strain during the first measurement and subsequently an increase in current with strain during the second measurement. Once the strain is released the second time, $I_D$ and $\mu_{FE}$ return to the same levels as before the second measurement. We point out that the device measurements displayed in Figure S9 and Figure S10 have higher initial unstrained mobilities and drive currents than the device measurements shown in Figure S6, but we were unable to further improve the mobility with strain because of device break-in and degraded electrical performance at 0.8% strain.

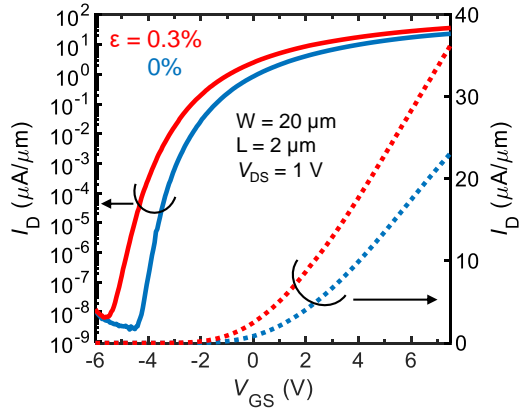

**Figure S11.** $I_D$-$V_{GS}$ measurements at 0% and 0.3% strain for a transistor with $W = 20$ μm and $L = 2$ μm on a different sample. The mobility increases from ~25 cm²V⁻¹s⁻¹ at 0% strain to ~34 cm²V⁻¹s⁻¹ at 0.3% strain at $n$ ~ 7×10¹² cm⁻².

We also perform strain-dependent electrical measurements on a third set of monolayer $MoS_2$ transistors with higher initial mobilities (~15 cm²V⁻¹s⁻¹ to 35 cm²V⁻¹s⁻¹). Transfer characteristics from one of these transistors are included in Figure S11. The mobility and current of these devices improve with ~0.3% strain, but the devices degrade at higher levels of strain (not shown). The mobility of the device in Figure S11 increases from ~25 cm²V⁻¹s⁻¹ with 0% strain to ~34 cm²V⁻¹s⁻¹ with 0.3% strain at $n$ ~ 7×10¹² cm⁻².

We attribute the degradation in electrical performance at higher strains to cracking in the metal or $Al_2O_3$ gate dielectric, or worsened adhesion of the contact metal to the $MoS_2$. When strain greater than 1% is applied to the $MoS_2$ devices, we observe obvious cracks in the metal or $Al_2O_3$ layers (see Figure S12). Although we did not observe such cracks in the devices measured in Figures S9-S11, this could still be the source of device degradation. Raman measurements (see Figure S13) showing the expected shift in the E' peak prove that the $MoS_2$ remains strained at 0.8%, and that the $MoS_2$ is not "slipping" against the substrate.

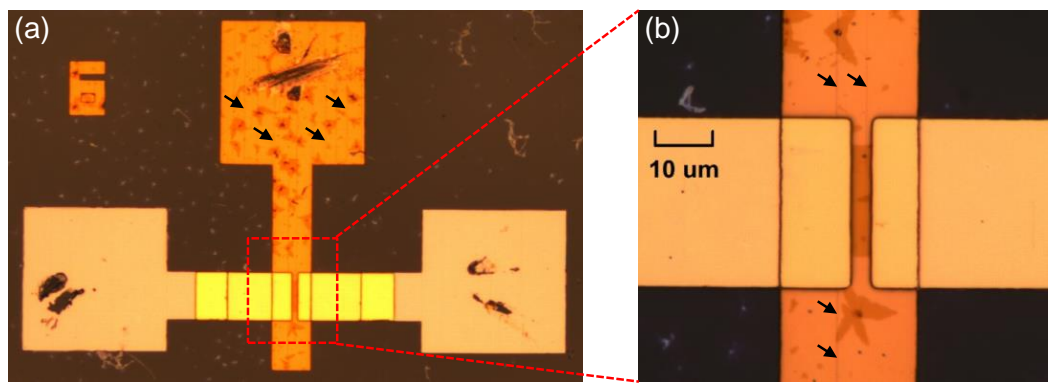

**Figure S12.** (a) Optical image of a device after >1% strain was applied. (b) Zoomed-in region showing fine cracks in the metal gate and/or the $Al_2O_3$ gate insulator under the $MoS_2$ channel. Small arrows point to the fine cracks.

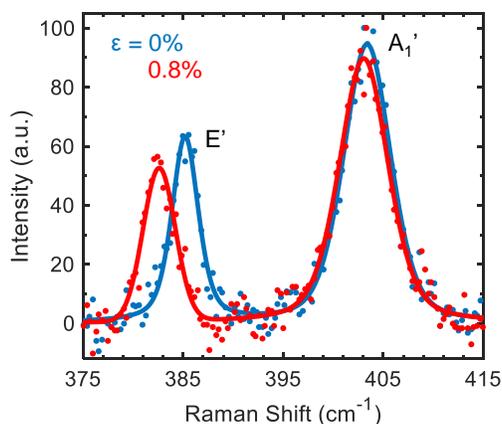

**Figure S13.** Raman spectra at 0% and 0.8% tensile strain for one of the devices that degraded electrically at 0.8% strain. The data are depicted as symbols, and the fits using a superposition of Lorentzian and Gaussian peaks are displayed as solid curves. The redshift in the E' peak shows that the $MoS_2$ is still strained (even after the electrical degradation of the transistor) and does not slip against the substrate.

# 7. Supplementary References:

(1) Münzenrieder, N.; Petti, L.; Zysset, C.; Görk, D.; Büthe, L.; Salvatore, G. A.; Tröster, G. Investigation of gate material ductility enables flexible a-IGZO TFTs bendable to a radius of 1.7 mm. *2013 Proceedings of the European Solid-State Device Research Conference (ESSDERC)*, 16-20 Sept. 2013, **2013.** 10.1109/ESSDERC.2013.6818893.

(2) Smithe, K. K. H.; English, C. D.; Suryavanshi, S. V.; Pop, E. Intrinsic electrical transport and performance projections of synthetic monolayer $MoS_2$ devices. *2D Materials* **2017,** *4*, 011009.

(3) Smithe, K. K. H.; Suryavanshi, S. V.; Muñoz Rojo, M.; Tedjarati, A. D.; Pop, E. Low Variability in Synthetic Monolayer $MoS_2$ Devices. *ACS Nano* **2017,** *11*, 8456-8463.

(4) Vaziri, S.; Yalon, E.; Munoz Rojo, M.; Suryavanshi, S. V.; Zhang, H.; McClellan, C. J.; Bailey, C. S.; Smithe, K. K. H.; Gabourie, A. J.; Chen, V.; Deshmukh, S.; Bendersky, L.; Davydov, A. V.; Pop, E. Ultrahigh thermal isolation across heterogeneously layered two-dimensional materials. *Sci Adv* **2019,** *5*, eaax1325.

(5) Buscema, M.; Steele, G. A.; van der Zant, H. S. J.; Castellanos-Gomez, A. The effect of the substrate on the Raman and photoluminescence emission of single-layer $MoS_2$. *Nano Research* **2014,** *7*, 561-571.

(6) Conley, H. J.; Wang, B.; Ziegler, J. I.; Haglund, R. F.; Pantelides, S. T.; Bolotin, K. I. Bandgap Engineering of Strained Monolayer and Bilayer $MoS_2$. *Nano Letters* **2013,** *13*, 3626-3630.

(7) Wang, Y.; Cong, C.; Qiu, C.; Yu, T. Raman Spectroscopy Study of Lattice Vibration and Crystallographic Orientation of Monolayer $MoS_2$ under Uniaxial Strain. *Small* **2013,** *9*, 2857-2861.

(8) Shi, H.; Pan, H.; Zhang, Y.-W.; Yakobson, B. I. Quasiparticle band structures and optical properties of strained monolayer $MoS_2$ and $WS_2$. *Physical Review B* **2013,** *87*, 155304.

(9) Goswami, T.; Rani, R.; Hazra, K. S.; Ghosh, H. N. Ultrafast Carrier Dynamics of the Exciton and Trion in $MoS_2$ Monolayers Followed by Dissociation Dynamics in $Au@MoS_2$ 2D Heterointerfaces. *The Journal of Physical Chemistry Letters* **2019,** *10*, 3057-3063.

(10) Mak, K. F.; Lee, C.; Hone, J.; Shan, J.; Heinz, T. F. Atomically Thin $MoS_2$: A New Direct-Gap Semiconductor. *Physical Review Letters* **2010,** *105*, 136805.

(11) Liu, Z.; Amani, M.; Najmaei, S.; Xu, Q.; Zou, X.; Zhou, W.; Yu, T.; Qiu, C.; Birdwell, A. G.; Crowne, F. J.; Vajtai, R.; Yakobson, B. I.; Xia, Z.; Dubey, M.; Ajayan, P. M.; Lou, J. Strain and structure heterogeneity in $MoS_2$ atomic layers grown by chemical vapour deposition. *Nature Communications* **2014,** *5*, 5246.

(12) Cai, L.; McClellan, C. J.; Koh, A. L.; Li, H.; Yalon, E.; Pop, E.; Zheng, X. Rapid Flame Synthesis of Atomically Thin $MoO_3$ down to Monolayer Thickness for Effective Hole Doping of $WSe_2$. *Nano Letters* **2017,** *17*, 3854-3861.

(13) Zhao, Y.; Xu, K.; Pan, F.; Zhou, C.; Zhou, F.; Chai, Y. Doping, Contact and Interface Engineering of Two-Dimensional Layered Transition Metal Dichalcogenides Transistors. *Advanced Functional Materials* **2017,** *27*, 1603484.

(14) Schauble, K.; Zakhidov, D.; Yalon, E.; Deshmukh, S.; Grady, R. W.; Cooley, K. A.; McClellan, C. J.; Vaziri, S.; Passarello, D.; Mohney, S. E.; Toney, M. F.; Sood, A. K.; Salleo, A.; Pop, E. Uncovering the Effects of Metal Contacts on Monolayer $MoS_2$. *ACS Nano* **2020,** *14*, 14798–14808.

(15) Datye, I. M.; Gabourie, A. J.; English, C. D.; Smithe, K. K. H.; McClellan, C. J.; Wang, N. C.; Pop, E. Reduction of hysteresis in $MoS_2$ transistors using pulsed voltage measurements. *2D Materials* **2019,** *6*, 011004.

(16) English, C. D.; Shine, G.; Dorgan, V. E.; Saraswat, K. C.; Pop, E. Improved Contacts to $MoS_2$ Transistors by Ultra-High Vacuum Metal Deposition. *Nano Letters* **2016,** *16*, 3824-3830.

(17) Hosseini, M.; Elahi, M.; Pourfath, M.; Esseni, D. Strain induced mobility modulation in single-layer $MoS_2$. *Journal of Physics D: Applied Physics* **2015,** *48*, 375104.